# A digital business ecosystem maturity model for personal service firms

Ricardo Guerrero, Christoph Lattemann, Simon Michalke and Dominik Siemon.

<a>INTRODUCTION

In recent years, companies worldwide have been facing significant and fundamental challenges due to increasing trends towards digitalization. Nowadays, the environment of organizations is changing faster and has become more volatile, uncertain, and complex than in the past. Rapid changes in a business ecosystem (BE) concerning aspects related to the strategy, business model, collaboration practices, solutions portfolio, technology, knowledge management, and communication, make it more important than ever for firms to be able to respond and adapt to their environments (Kaufman and Horton, 2015; Schuchmann and Seufert, 2015). In this context, firms from all industries must seek appropriate working modes to become more agile, build up dynamic capabilities, and search for adequate organizational forms that enable innovation and value creation at unprecedented speed, scale, and impact (Gorissen et al. 2016).

In the past years, the industrial production and manufacturing sector, as well as few service sectors such as financial, logistics, healthcare, or telecommunications, have found ways to develop digital business ecosystems (DBEs) by applying concepts of Industry 4.0 (Ibarra et al. 2018; Matt et al. 2020; Valdez-de-Leon, 2016). For instance, firms such as Tesla, Rolls-Royce, and Volkswagen serve as examples in the manufacturing sector (Ng et al. 2012; Smolnicki and Sołtys, 2016). In the logistics and finance sector, companies such as DHL, Hermes, N26, and TransferWise, also serve as examples (Kersten et al. 2017). Consequently, firms such as Huawei and Zava are tremendous examples from the telecommunication and healthcare sector (Hermes et al. 2020; Tao and Chunbo, 2014). Although many firms from multiple sectors have succeeded in establishing DBEs, the so-called "personal services" (PS) (Larsson, 2015; Lattemann et al. 2019), represent one of the few service sectors struggling the most to achieve the transition from analogue to digital and thus, primarily relying on traditional and non-digital BEs.

PS are services that take place at the human being as the service object and are characterized by having high-contact levels of interaction, being usually co-created, and tailor individual needs and desires (Lattemann et al. 2019). PS aim to stabilize or improve a human being's life situation. They are essential for society and everyday life. These services can be found in sectors such as education, retail, hospitality (hotels, restaurants), and craftsmanship (Guerrero et al. 2020; Mattila and Enz, 2002; Parasuraman et al. 1985). For so long, PS have been purely relying on analogue processes, front-desk and face-to-face activities (Guerrero et al. 2020; Lattemann et al. 2020). This has led these kinds of firms to suffer extreme consequences and challenges, especially during the coronavirus (COVID-19) pandemic times, wherein many countries have imposed lockdowns and social distancing norms (Agarwal et al. 2020; Bartik et al. 2020). In this context, such as in other industries, PS must also go through a digitalization process allowing these firms to build DBEs that involves the application of information and communication technology (ICT).

Digitalization can make an important and supportive contribution to PS and change its interaction space within all its facets (Lattemann et al. 2020). However, what many of these firms lack today are strategic instruments that allow them to implement adequate processes and practices to effectively manage and guide the transition towards designing a DBE systematically (Larsson, 2015). An instrument that allows firms to identify gaps for improvements concerning the development of DBEs is a maturity model (MM) (Becker et al. 2009; Cukier and Kon, 2018; Jansen, 2020). MMs help assess firms' current developmental (digitalization) stage and show trajectories to guide the transition towards a DBE in a well-structured way (Teichert, 2019).

Several MMs have been proposed for various business sectors (e.g., software development, manufacturing, public services), allowing these firms to adequately develop DBEs (Jansen, 2020; Pullen, 2007). However, none of these MMs are applicable nor specific to PS and, as such, do not offer specific guidance for this sector. To address this gap, this research aims at developing an MM that focuses on providing specific and targeted initiatives for improvement towards the development of a DBE. In this paper, we strive to answer the following research questions: (1) *Which function, process, or capability area need to be considered when defining a DBE for PS firms?* (2) *What are the maturity stages needed to develop a DBE within PS firms?* To address these questions, we follow a Design Science Research (DSR) approach (Hevner et al. 2004).

The remainder of the paper is structured as follows. Firstly, we present the theoretical background on the digitalization of PS. Secondly, we discuss research on DBE and MM. Thirdly, we discuss research on the Service-Dominant Logic (S-D logic). Fourthly, our applied research method is introduced and our interview results are presented and discussed. Finally, the paper ends with a discussion of the results and conclusions.

<a>THEORETICAL BACKGROUND

This section provides an overview of core concepts and presents how PS firms and DBE research streams are interconnected.

<b>Digitalization of Personal Services

PS are "services including high-contact levels of interaction" (Kellogg and Chase, 1995, p. 1737) with the customers. They are characterized by intimacy, the exchange of content-rich information, and long interaction times. They encompass feelings and emotional states such as friendliness, helpfulness, and empathy (King and Garey, 1997; Parasuraman et al. 1985) and are developed in a co-created manner (Lattemann et al. 2020). PS are all about the fulfillment of individual, human needs in the situation of "using" the service - the so-called "Value-in-Use" (Grönroos, 2008; Vargo and Akaka, 2009) - and also about the fulfillment of user experiences in interaction - the so-called "Value-in-Interaction" (Geiger et al. 2020). Here, both (*Value-in-Use* and *Value-in-Interaction)* could be highly influenced by ICT. According to Lattemann et al. (2020), through the application and dynamic developments of ICT, the nature and exchange of PS undergo radical shifts, affecting the design and provision of such services (e.g. video-streaming selling instead of front-line selling). ICT results in opening new perspectives of PS regarding value, communication, interaction, collaboration, and co-creation, which are fundamental aspects when developing a DBE (Robra-Bissantz et al. 2020).

Finally, according to a study conducted by Bartik et al. (2020) in an attempt to explore the impact of COVID-19 across multiple PS firms (i.e. retailers, arts and entertainment, and hospitality) in the United States of America (USA), results revealed that among a sample of more than 5.800 businesses, 43 percent of them had temporarily closed. Nearly all of these closures were due to COVID-19. Consequently, their results also revealed that if by December 2021, the COVID-19 regulations such as lockdowns and social distancing norms imposed by the USA continue, 53 percent of such businesses could lead to bankruptcy. Therefore, they concluded that as other business industries and especially due to COVID-19 strict social distancing norms (e.g. minimizing personal interaction at the very least), PS firms must find ways to design and develop their services relying on the application of ICT.

<b>Digital Business Ecosystems

BEs have been continually defined, re-defined and studied over the past three decades (Moore, 1993; Selander et al. 2010). Inspired by the logic of biological ecosystems, James F. Moore introduced the metaphoric concept of "business ecosystem" in 1993 (Moore, 1993). In his work, BEs are defined as "an economic community supported by a foundation of interacting organizations and individuals – the organisms of the business world" (Moore, 1993; p. 76). This economic community produces goods and services of value to customers, who themselves are members of the ecosystem. Nonetheless, the advancements in ICT have led to the development of new collaborative organizational networks such as DBEs (Senyo et al. 2019). DBEs extend the concept of BEs by emphasizing ICT. DBEs transcend traditional industry boundaries to foster open and flexible collaboration and competition (Senyo et al. 2019).

Scholarly work differs significantly in the definition and interpretation of DBEs (Razavi et al. 2010; Senyo et al. 2019). Building on a purely technical perspective, Nachira et al. (2007) define DBEs as "a virtual environment populated by digital entities such as software applications, hardware and processes" (Nachira et al. 2007, p. 9). A broader perspective incorporates the view of network theory, which assesses the underlying issues of exchange processes that take place through interaction between entities and a number of individual actors within a digital ecosystem (Camarinha-Matos and Afsarmanesh, 2008). Camarinha-Matos and Afsarmanesh (2008) define DBEs as "a class of collaborative networks with a wider alliance of heterogeneous and geographically dispersed entities that collaborate via the Internet to achieve common outcomes" (Camarinha-Matos & Afsarmanesh, 2008; p. 62). Another, even broader perspective focuses on the S-D logic and lays the focus on value co-creation (Senyo et al. 2019; Vargo and Lusch, 2004). According to Senyo et al. (2019), DBEs can be defined as "socio-technical environments made up of different individuals and organizations with collaborative and competitive relationships to collectively co-create value through ICT and the coordination of superb management practices" (Senyo et al. 2019, p. 125). For the purpose of this paper, we follow the definition established by Senyo et al. (2019) and argue that co-created value is presumed to be greater than that created by a single organization (Adner, 2006). Consequently, other authors such as Breidbach and Maglio (2016) have revealed that ICT can be used to transform and enhance value co-creation processes. Thus, given the fact that DBEs rely on synergies between different actors to generate value, we see value co-creation through the use and application of ICT as an important driver in DBEs formation and operation.

**Service-Dominant Logic**

The S-D logic is rooted in marketing research, where it gained momentum since its inception by the landmark study of Vargo and Lusch (2004), followed by further amendments (Vargo and Lusch, 2016, 2008). The S-D logic is one of the most relevant service theories that, based on network theory assumption, conceptualizes markets as networks of co-creating actors (Vargo, 2009; Vargo et al. 2008). Based on Akaka and Vargo (2014), the main considerations of S-D logic are the following: "(1) services is the basis of exchange, (2) value is always co-created among multiple stakeholders in a business ecosystem, (3) all social and economic actors are resource integrators, and (4) value is always contextually and phenomenologically derived" (Akaka and Vargo, 2014, p. 4). The S-D logic contradicts other marketing theoretical frameworks such as the so-called "Goods-Dominant Logic" (G-D logic), which is based on the assumption that the producer and customer are strictly separated from each other and that the value of the tangible asset or product is defined by the market price or what the client is willing to pay (Value-in-Exchange) (Jallat, 2004). Under the S-D logic, companies cannot create value by themselves but rather focus on the cooperation of different actors (e.g. service providers, customers) and stakeholders to apply collective knowledge to create value for the individual and the entire business ecosystem (co-creation) (Vargo and Lusch, 2004). In this view, customers and other stakeholders are increasingly involved in the process of service delivery and all of them contribute to the creation of value.

Additionally, from an S-D logic perspective, ICT itself is considered as an independent actor as well as a resource for value creation (Lusch and Nambisan, 2015). In this context, by viewing ICT as an actor, other actors can extend their ability to reconfigure resource integration within the ecosystem itself, such as IT capability, and knowledge sharing and coordination (Nambisan, 2013). Thus, often leading to the design and development of DBEs (Senyo et al. 2019). This highlights how the fundamental function of co-creation, the peculiar view towards ICT, and the inherent ecosystem perspective, makes the S-D logic a natural fit to be taken into consideration when designing and developing DBEs (Senyo et al. 2019; Sklyar et al. 2019).

<b>Maturity Models

MMs are strategic frameworks that allow firms to assess internal and external improvement processes, enabling a transformation from ad hoc process implementation to definitive and disciplined process execution (Becker et al. 2009; Bruin et al. 2005). The main purpose of MMs is to provide specific and targeted improvement initiatives towards a specific business function, process, or capability area. They help firms to encounter digitalization according to predefined dimensions (Schäffer et al. 2018). Especially in the case of the transition towards DBEs, MMs can assist in understanding the current state of 'maturity' and the capabilities of an organization in effectively managing and guiding digitalization efforts in a systematic way. Thus, leading to a stage of 'digital maturity' (Teichert, 2019). Here, digital maturity goes beyond a mere technological interpretation and also reflects a managerial understanding describing changes in the BE. Thus, digital maturity comprises technological and managerial aspects. The underlying structure of MMs is always similar: Firstly, critical capabilities are defined for a specific area of action (e.g. strategy, collaboration, knowledge management). Secondly, improvement initiatives are proposed for the capabilities defined at each of the different maturity stages, whereby the lowest stage stands for an initial state that can be characterized by an organization having little capabilities in the domain under consideration. In contrast, the highest stage represents a conception of high maturity (Becker et al. 2009). Finally, depending on which requirements are fulfilled concerning the different 'maturity stages', a certain degree of maturity is awarded (Schäffer et al. 2018).

<a>METHODOLOGICAL APPROACH

<b>Design Science Research (DSR)

To develop our MM, we followed the MM development process proposed by de Bruin et al. (2005) since this provides a well-defined process that is based on DSR guidelines (Hevner et al. 2004). DSR strives to build and evaluate 'artifacts' that are to be understood as constructs, models, guidelines, methods, or instantiations to solve organizational problems (Hevner et al. 2004). From a DSR perspective, MMs can serve as reference models and hence, are artifacts that show "an anticipated, desired, or typical evolution path" (Becker et al. 2009, p. 213). The development process is based on six phases: (1) *scope*, (2) *design*, (3) *populate*, (4) *test*, (5) *deploy*, and (6) *maintain*. Phases 1 to 3 are crucial to develop the design specifications of a MM, whereas phases 4 to 6 concern its application, evaluation, and long-term refinement. The research described in this paper comprises the development, application, and evaluation of the model (i.e. scope, design, populate, test, and deploy), while subsequent phases (maintain) are planned for future research, as this must be done on a long-term basis. Finally, for the development of our MM, initially, a literature review (LR) in accordance with Webster and Watson (2002) was conducted, which at the same time, led to the development of an initial MM. Once conducted the LR, our initial MM was further refined and validated through expert interviews and later on, tested by two PS firms' representatives to address issues of reliability and validity of our MM. Table 1 depicts our methodological approach in detail.

| Phase 1 - Scope | Building a MM can assist PS firms to understand their current 'digitalization' state and to provide specific and targeted improvement initiatives towards the development of DBEs. |
|---|---|
| Phase 2 - Design | A LR in accordance with Webster and Watson (2002) was conducted. The focus was set on literature related to MMs and DBEs as well as digitalization and PS. Web of Science was used as a database. A total of fifteen articles were selected and analyzed in-depth. Once analyzed the content of the fifteen articles, it became apparent that none of the MMs were applicable nor specific to PS. The initial MM was set up, indicating the identified stages, dimensions, capability areas, and improvement initiatives for some of the capability areas at each of the |
| | different stages. Capability areas (e.g. 'customer satisfaction', 'customer interaction', 'customers empathy', and 'product/service individualization') have never been proposed before in previous DBE nor digitalization MMs. We define their initiatives for improvement in phase 3 (populate) through the conduction of interviews. |
| **Phase 3 - Populate** | The initial MM was discussed with experts. The focus was laid on interviewing either founders, CEOs, firm's managers, and service/product designers from PS all over the globe (see table 2). Our study is based on a convenience sample: Interviewees are experts in the field of PS, digitalization, and BE. Consequently, interviewees were selected based on the authors´ network. A semi-structured interview guideline was conceptualized and a total of eleven semi-structured interviews of approximately 45 - 60 minutes were conducted. Here, questions to the interviewers were asked regarding whether they would consider the stages, dimensions, and capability areas firstly identified, relevant or not. The experts were also offered the opportunity to provide feedback to the initial architecture of the MM in ways that they could add, modify, or remove any of the previously identified stages, dimensions, capability areas, as well as improvement initiatives. The expert interviews were recorded, transcribed, and subsequently coded using MAXQDA as a computer-based qualitative analysis tool. The transcribed data were independently analyzed by two researchers, using codes as an efficient data-labelling and data-retrieval method. |
| **Phase 4 and 5 - Test and Deploy** | The MM was initially refined and validated through the conduction of semi-structured interviews, and then it was further tested with two PS firms from our interview network to gain more practical insights. |

Table 1: Design science research approach to develop our final maturity model

| *Company (Focus)* | *Position* |
|---|---|
| **Business as Unusual** (Retailing) | Co-Founder |
| **Business as Unusual** (Retailing) | Founder and CEO |
| **Neurosales** (Consulting) | Founder and CEO |
| **Clarke & Partners** (Event Management and Coordination) | Commercial Manager |
| **Pg40 Consulting Group** (Consulting) | Designer |
| **Endobrand** (Communication Agency) | CEO |
| **Friotem, C.A.** (Retailing) | Founder and CEO |
| **Boston Consulting Group** (Consulting) | Product/Service Designer |
| **Neurosales** (Consulting) | Service Manager Director |
| **Business as Unusual** (Retailing) | Chief Transformation Officer |
| **SOHICA, C.A.** (Healthcare) | Founder and CEO |

Table 2. Interview partners

## Literature Review

### Pivotal capabilities of a PS DBE

Through the analysis of existing MMs in DBE and digitalization (Alves et al. 2011; Azevedo and Santiago, 2019; Gollhardt et al. 2020; Schumacher et al. 2016; Valdez-de-Leon, 2016), common capabilities could be identified, which at the same time could be applied in a PS context. This process resulted in the derivation of over thirty capabilities. In order to create a holistic perspective and to identify critical capabilities applicable in the context of a PS DBE, we followed the approach established by Fraser et al. (2002). They recommend separating MM capabilities in a multi-dimensional manner and discard those capabilities that do not directly have an impact in the domain under consideration (i.e.

PS). This process resulted in selecting eight capabilities ('vision & strategy', 'business model', 'digital culture', 'knowledge management', 'collaboration', 'agility & flexibility', 'ICT-Infrastructure', and 'ICT-System application'). Further, through an analysis of the PS literature (Lattemann et al. 2020, 2019; Mattila and Enz, 2002; Parasuraman et al. 1985), we identified four unique capabilities of PS ('customer satisfaction', 'customer interaction', 'customers empathy', and 'product/service individualization'). In total, twelve capabilities were selected for the development of our initial MM. Finally, based on Fraser et al. (2002), these were further clustered and grouped into five dimensions (customer centricity, strategy, products/services, process & organization, and technology). The dimensions are described in the following.

(1) Customer-Centricity - measures the company´s ability to use ICT to collaborate with their customers and other actors in the BE. It includes concepts such as co-design and co-creation. Here, customers are considered as resources for value creation. It is analyzed by means of the following capabilities: *customer satisfaction*, *customer interaction*, and *customer empathy*.

(2) Strategy - measures the company´s ability to develop a strategic business/IT alignment plan and effectively implement it across all levels of the organization in such a way that a DBE could be guaranteed. It is analyzed by means of the following capabilities: *vision and strategy*, *business model*, and *culture*.

(3) Products/Services - measures the company´s ability to design products/services through collective efforts by a collaborative network, as well as the ability to integrate different technologies in the development process. It is analyzed by means of the following capability: *product/service individualization*.

(4) Process & Organization - measures the company´s ability to implement technological, management, and agile management practices to improve the BE. It is analyzed by means of the following capabilities: *knowledge management*, *collaboration*, and *agility and flexibility*.

(5) Technology - measures the company´s ability to understand which technologies are becoming relevant and could influence the BE. Here, companies determine which suitable ICT-systems are needed to support their employees in carrying out their tasks. It is analyzed by means of the following capabilities: *infrastructure* and *ICT-System application*.

### Maturity stages and their characteristics

We decided to follow the maturity stages proposed by Teichert (2019), as he built on the concept of 'digital maturity' and we considered him to provide the most accurate and solid view towards achieving a DBE. In this context, we classified the different maturity stages into: (1) infancy, (2) developing, (3) transforming, (4) optimized, and (5) digital maturity.

The stages are described in the following.

*Stage 1 - Infancy.* The organization has low to zero awareness of the benefits of using ICT and how to build a DBE. The initiatives towards innovation in products/services are poorly addressed. The firm relies purely on analog and undisciplined processes.

*Stage 2 - Developing.* Organizations recognize the need to implement ICT to improve their BE. However, ICT is still inconsistent with low strategy implementation towards how a DBE should be achieved. Most of the firm's operation processes are still analog.

*Stage 3 - Transforming.* Basic ICT-Systems are presented in order to improve collaboration and interaction among customers and other BE actors. Some processes become digital; however, they are not totally under control. Therefore, the outcomes are not predictable. The firm has a more sophisticated management structure, which encourages employees to learn and use ICT-Systems.

*Stage 4 - Optimized.* The application and use of ICT is reflected in the firm's strategy. The organization's processes are well understood, and the responsibilities and roles of all BE actors are clear and well-defined. High levels of collaboration and coordination is achieved through the use of ICT. Analog business functions disappear and processes take shape through deeply embedded ICT-Systems.

*Stage 5 - Digital Maturity.* The firm can continually improve its BE with incremental or radical innovation, technologies, and resources. Concrete strategy to improve service experience through a complete understanding of the customers and use of ICT.

Table 3 describes our initial MM derived from the LR.

| Stages | Dimensions | Capabilities | Initiatives for Improvement |
|---|---|---|---|
| Stage 1 - **Infancy** Stage 2 - **Developing** Stage 3 - **Transforming** Stage 4 - **Optimized** Stage 5 - **Digital Maturity** (Teichert, 2019) | Customer Centricity | *Customer Satisfaction* (King and Garey, 1997; Mattila and Enz, 2002; Parasuraman et al. 1985) | - |
| | | *Customer Interaction* (King and Garey, 1997; Lattemann et al. 2020, 2019; Parasuraman et al. 1985; Robra-Bissantz et al. 2020) | - |
| | | *Customer Empathy* (King and Garey, 1997; Mattila and Enz, 2002; Parasuraman et al. 1985; Vink, 2018) | - |
| | Strategy | *Vision & Strategy* (Canetta et al. 2018; Gollhardt et al. 2020; Valdez-de-Leon, 2016) | Vision & Strategy (stage 1 - 5) |
| | | *Business Model* (Gollhardt et al. 2020; Lattemann et al. 2020, 2019; Robra-Bissantz et al. 2020; Schumacher et al. 2016; Teichert, 2019) | Business Model (stage 1 - 5) |
| | | *Digital Culture* (Azevedo and Santiago, 2019; Gollhardt et al. 2020; Teichert, 2019) | Digital Culture (stage 1 - 5) |
| | Product and Services | *Product/Service Individualization* (Lattemann et al. 2020, 2019; Lindh and Nordman, 2018) | - |
| | Process and Organization | *Knowledge Management* (Gollhardt et al. 2020; Schumacher et al. 2016; Teichert, 2019; Valdez-de-Leon, 2016) | Knowledge Management (stage 1 - 5) |
| | | *Collaboration* (Lattemann et al. 2020, 2019; Schumacher et al. 2016; Teichert, 2019) | Collaboration (stage 1 - 5) |
| | | *Agility and Flexibility* (Gunsberg et al. 2018; Teichert, 2019; Valdez-de-Leon, 2016) | Agility and Flexibility (stage 1 - 5) |
| | Technology | *Infrastructure* (Carvalho et al. 2019; Lattemann et al. 2020, 2019; Teichert, 2019) | Infrastructure (stage 1 - 5) |
| | | *ICT-System Application* (Carvalho et al. 2019; Lattemann et al. 2020, 2019; Teichert, 2019) | ICT-System Application (stage 1 - 5) |

Table 3: Initial design of the maturity model derived from literature

<a>RESULTS

<b>Validation and Refinement of the Initial MM through Semi-Structured Interviews

<c>Interview results concerning the pivotal capabilities of a PS DBE

All interviewees recognized our capability areas identified to be suitable to measure a PS DBE. Nonetheless, six out of eleven interviewees expressed the necessity to include 'leadership' as a key aspect of our MM. In this context, expert 2 commented: *"For us to be able to change our analog working modes towards more digital and dynamic ones, a structured form of leadership is required. PS firms need leaders that allow them to achieve a solid business/IT alignment and promote ICT-Systems (i.e. digital platforms, enterprise resource management systems, cloud-based services). These types of ICTs enhance management practices in ways that promote aspects concerning collaboration and co-creation […]"*. In the same lines, expert 7 commented: *"Leadership has to do a lot with encouraging your employees to attain your business goals and objectives but it also plays a pivotal role in defining a digital business model as well as promoting a digital culture, which are essential elements of a DBE […]"*. For instance, the initial dimension '*strategy*' was renamed into '*leadership & strategy*' to achieve a better fit and understanding towards our MM. Consequently, all interviewees expressed that the only possibility to assess a stage of digital maturity with regards to '*customer satisfaction*' and '*customer interaction*' was by implementing trend technologies such as artificial intelligence (AI), augmented reality (AR), chatbots, machine learning (ML), as well as by leveraging your business to an e-commerce level (i.e. building an integrated platform). Expert 1 commented: *"Making use of technologies such as AI and ML could accelerate your business in ways that you reach new standards to analyze customers' data. Data is the fuel that makes the whole customer journey works. AI and ML allow you to track all customer interactions (i.e. touchpoints), and thus, you are able to know what a customer wants at any given time. This will allow you to design much more individualized solutions […]"*. Expert 8 also commented: *"We were mostly a firm that managed everything on an analog basis. However, when COVID-19 arrived, everything changed for us. We had to close our business for a while and we were not able to sell anything. We were forced to leverage our business into an e-commerce scale by selling our products on a digital platform […]"*. Consequently, there was a consensus among the interviewees expressing that although the use of ICT can increasingly influence relationships between different actors in the BE, an empathic connection can be put at risk. Thus, a stage of digital maturity concerning '*customers empathy*' does not necessarily resemble the application of trend technologies but rather the action of favorable management practices by having the customer always in focus. Expert 3 commented: *"Our customers are one of the most and, if not, the most important actors from our BE. However, it is quite difficult to think of ways on how ICT might help customers' empathy. I feel like humans can only empathize with other humans. However, if you promote your firm's employees to reach an emotional connection with their customers and adopt a customer-centered perspective at all times, I doubt that you will end up losing customers […]"*. Similarly, when discussing aspects such as '*products/service individualization*', '*knowledge management*', '*collaboration*', '*agility and flexibility*', '*ICT-Infrastructure*', and '*ICT-Systems application*', there was a consensus among the interviewees addressing that in order to reach a digital maturity stage in such capability areas, the application of trend technologies were extremely important. However, the assessment of one and the other varied depending on each case. For instance, in the case of '*product/service individualization*', six out of eleven interviewees claimed that reaching a stage of digital maturity depended more on the type of technologies applied when developing a product/service. Expert 11 commented: *"The first step for successful individualization is to gather comprehensive and accurate data sets. This, you can do by making use of technologies such as AI and ML. If you become an expert on these techniques, you will be able to track and improve users' experiences at any time. Thus, you will be able to always provide value to your customers anytime […]"*. In the context of '*knowledge management*' and '*collaboration*', all interviewees claimed that reaching a stage of digital maturity depended more on the type of technologies applied to improve management practices. Expert 4 commented: *"Having a digital platform allows us to enhance our BE to another level. Aspects such as time and space are not relevant anymore. With our digital platform, multiple and different actors are able to interact with each other at all times. We have customers in the USA, Latin America as well as in Europe, and it is often the case that you can see them communicating with each other. It gives not only us as a firm but also them another sense of collaboration […]"*. In the context of '*ICT-Infrastructure*' and '*ICT-Systems application*', all interviewees claimed that reaching a stage of digital maturity depended more on the type of technologies built on the ICT-Systems as well as infrastructure of the firm. Expert 2 commented: *"After COVID-19, analog functions and paper-based processes are simply not possible anymore. For many of us, even personal interaction as well as in-door selling is not possible anymore. Having ICT tools such as Zoom,*

*Skype, or any other ICT-Systems are the new means to collaborate with our customers and partners. In fact, these ICT tools are the least you can have. Many businesses are right now using technologies such as virtual agents, digital platforms, etc., which gives them another sense of collaboration and interaction with the different actors of their BE […]"*. Finally, when discussing the aspect related to '*agility and flexibility*', six out of eleven interviewees emphasized the use of agile methods (i.e. SCRUM and Design Thinking) as activities to enhance a DBE due to the fact that these were management practices that allowed firms to achieve new kinds of innovations with regards to products or services as well as to collaborate with their customers and other actors from the BE. Thus, enabling new forms of co-creation. They also indicated that the execution of these practices through the implementation of ICT tools (e.g. digital whiteboards) could represent a digital maturity stage.

### <c>Interview results concerning the maturity stages and their characteristics

The initial maturity stages derived from the LR were discussed with the experts. The discussion revealed that although Teichert (2019) built on the concept of DBE by proposing its view on 'digital maturity', the characteristics of such stages were too narrow, as they implied a more 'general' perspective rather than being industry-specific and being applicable for the PS sector. As a result, we suggest a fit between the characteristics provided by Teichert (2019) and our findings gathered from our interviews. In the following, the maturity stages and their descriptions are illustrated in table 4.

| **Stage 1 - Infancy.** | PS firm relies purely on analog and undisciplined processes. Business/IT alignment is not achieved or aimed to improve BE but instead, the firm tries to reduce costs as much as possible and sees ICT as a high-cost unit center. Ineffective management decisions are taken. The firm is short-term focused. Co-creation is unknown. |
|---|---|
| **Stage 2 - Developing** | PS firm begins to understand the application and use of ICT as a necessity to improve BE. Nonetheless, most firms' operation processes are still analog. Business/IT alignment is not achieved rather characterized by informal and limited employees' heroic actions. Outputs and customer involvement are inconsistent but the firm begins to recognize that focusing on the customer might enhance the BE and develop innovative service/products outputs. |
| **Stage 3 - Transforming.** | PS firm has understood the application and use of ICT as a necessity to improve BE. The firm has defined plans and strategies on how to achieve business/IT alignment in such a way that it has a formal and omnipresent relation. Management is more sophisticated, open, and engaged towards applying trend technologies (i.e. AI, AR, virtual agents, digital platforms). The firm has its first co-creative efforts to improve collaboration and interaction aspects among different actors in the BE. |
| **Stage 4 - Optimized.** | The application and use of ICT is reflected in the firm's strategy. PS firm applies trend technologies (see. stage 3) to improve aspects related to interaction, collaboration, co-creation, and thus, enabling new relationship settings among all different actors involved in the BE. Business/IT alignment is achieved, leading to well-prioritized digital projects and engagement in the strategy development process. Customers and all other actors in the BE are considered as big sources of value contribution. Innovative outputs are visible and acknowledged due to co-creation and co-design practices. |
| **Stage 5 - Digital Maturity.** | PS firm goes beyond business/IT alignments, and formulate and execute an organization strategy by leveraging digital resources to create differential value. The PS firm relies mainly on the use and application of trend technologies to improve aspects related to interaction, collaboration, and co-creation, leading to extraordinary relationships among all different actors involved in the BE. The firm has a coherent digital business strategy, which is well communicated throughout the organization and should be treated as a business strategy in the digital landscape. Co-creation is embedded in a firm's mindset and culture. |

Table 4: Revised maturity stages and their descriptions.

### Validation of the revised MM through real-world application

We further evaluated the usefulness and applicability of our MM in a real-world scenario with two PS companies from our interview network. Both companies are on the road to digitalization and are looking for ways for improvements on how to achieve a DBE. The feedback to our MM was overall positive in terms that the model showed a 'coherent' path on how PS firms could lead to achieve a DBE. However, both interviewees emphasized that the model should be considered as 'industry-specific' rather than 'context-specific', as it is obvious that not all PS firms are equal nor have similar behaviors. Company A interviewee commented: *"The MM could be more precise. However, for this to happen, you would have to adapt the MM to the requirements and characteristics of a specific firm. This means, the MM would be then 'context-specific'. However, I see the advantages of the model being applied on any PS firm embarking its way towards digitalization and achieving a DBE"*. Consequently, they also emphasized that just because a company does not possess a stage of total 'digital maturity' in all capability areas, it does not mean that the company has not achieved a DBE. In these lines, both interviewees claimed that reaching a stage of digital maturity highly depends on the size as well as resource availability and capacity of the firm. However, a DBE was considered to be achieved when the company has reached a certain business/IT alignment, which led the company to have fruitful and notable improvements and outputs. Company B interviewee commented: *"I understand that there are always new ways on how to make your BE more digital than it already is. This might mean relying, mainly, on the application of trend technologies (e.g. AI, AR, virtual agents, digital platforms). However, once the company achieved a proper balance of good ICT-Systems and practices with favorable management practices (i.e. co-creation), we consider having a DBE as well"*.

The final DBE MM for PS firms is presented in table 5.

| Dimension | Capability Area | 1. Infancy | 2. Developing | 3. Transforming | 4. Optimized | 5. Digital Maturity |
|---|---|---|---|---|---|---|
| Products & Services | Product/Service Individualization | Firm considers product/service individualization to be of no priority and management sees no need to find ways to achieve it. Firm earns no income from the sale of individualized products/services. | Small initiatives towards individualized products/services start to appear. In this context, the firm, in an analog manner, inspects aspects related to the customers buying records, and conducts surveys to determine customers' behavior and needs. Firm recognizes the importance of investing on ICT to better develop individualized products/services but has neither | Groups of individualized products/services become more visible due to high efforts on analog practices as well as by implementing customer-relationship management (CRM) tools (e.g. GoogleTrends, Google Analytics, Social Media Analytics) and visual configuration softwares (e.g. Axonom, Powertrak). Firms' income start to become tangible due to the sale of | Ability to design and develop individualized products/services is enhanced by applying trend technologies (e.g. ML, AI, programmable robotic systems, crowdsourcing, 3-D printing). A high amount of firms' income is achieved through the development of individualized products/services. | Firm is an expert on designing and developing individualized products/services and relies on applying trend technologies such as (e.g. ML, AI, programmable robotic systems, crowdsourcing, 3-D printing). The majority of firms' income is achieved through the development of individualized products/services. |
| | | | the financial resources nor the know-how to put them into practice. | individualized products/services. | | |
| **Process & Organization** | | Knowledge Management | No proper knowledge transfer, creation, sharing, and application. Firm has no intention to manage organizational knowledge. | Management recognizes that knowledge management may be of value but is unwilling to search for ways to improve it. Exchange of knowledge is based, purely, on analog practices, and it happens within the organization but has no power to reach external actors. | Management is aware of the influence of ICT on knowledge sharing. Firm starts to adopt internet-based applications (e.g. social media platforms), enterprise resource management systems (e.g. SAP, Oracle), and document management softwares (e.g. Google Drive, Dropbox) to foster the exchange of knowledge at all times. | Knowledge management is deeply combined through the application of ICT tools to foster knowledge sharing at all times inside and outside of the organization. ICT tools for knowledge management and knowledge sharing are utterly established and acceptably used by the organization. | Knowledge management processes are reviewed and improved regularly. Knowledge management can be easily adapted to new organization's needs. Integrated platforms (e.g. digital platforms) are used for customers and other users to co-create and exchange knowledge at all times. |
| | | Collaboration | Collaboration takes place purely within functional silos. Departmental thinking is pronounced within the firm. | Collaboration still takes place, mostly, in functional silos. However, firm recognizes the importance of investing on ICT tools to improve communication and find better ways of collaboration. | Internet-based applications (e.g. social media platforms), document management softwares (e.g. Google Drive, Dropbox), and interactive whiteboards (e.g. Mural) are used to optimize internal as well as external collaboration. | Collaboration is mastered by applying and using trend technologies (e.g. integrated platforms, cloud-based services, chatbots), leading the firm to have co-created outputs. | Collaboration is fully digitalized through the application and use of trend technologies (e.g. integrated platforms, cloud-based services, chatbots), and the firm remains proactive in finding new ways for improvements. |

| | | | | | | |
|---|---|---|---|---|---|---|
| | Agility and Flexibility | Agile actions are principally unknown, and the technological basis is fragmented and unable to support agile processes effectively. Organizational activities for improving collaboration and cooperation do not take place. | Agile actions and technological implementation are, partly, implemented in some but not the majority of departments, business areas, teams, or structural levels of the organization. Minority of employees share agile competencies regarding communication, learning, responsibility, and customer-orientation. Only some employees are able to manage change appropriately. | Operational as well as technological changes are welcomed and handled accordingly. In many instances, the firm carries out activities to support and promote teamwork and establishes organizational structures that are flexible enough to cope with upcoming changes. | Firm invests on executive training for employees and managers to learn agile practices (e.g. SCRUM and Design Thinking). Firm manages to establish a proper technological basis throughout the entire organization and agile values (i.e. user-centeredness, openness to collaboration, acceptance to uncertainty, and openness to new risks) are shared and accepted among firms' employees. | Firm practices agile methods (e.g. SCRUM and Design Thinking) on a digital scale by making use of ICT tools (e.g. interactive digital whiteboards) to achieve firm's innovations as well as to improve management practices (i.e. collaboration, co-creation). All employees and managers have the competencies to work in an agile and changing environment successfully. Collaboration and cooperation are important aspects of everyday work. Operational as well as ICT-Infrastructure is flexible enough to react to upcoming changes quickly. |
| **Technology** | Infrastructure | Firm relies on paper-based systems. Communication and collaboration takes place by using only text-technologies (e.g. fax, e-mails) as well as via telephone. Information sharing within the organization | Firm starts to use basic ICT-Softwares (e.g. document-management softwares (Microsoft 365), enterprise resource management systems (SAP, Oracle), communication and messaging | Firm extends basic ICT-Softwares through the application of Internet-based applications (e.g. social media platforms) as well as by including video-conference services (e.g. | Firm extends ICT-Infrastructure through the use and application of ICT-Systems (e.g. intelligent call routing, interactive voice responses, virtual receptionist). | ICT-Infrastructure is fully supported by implementing trend technologies (e.g. real-time messaging, AI, chatbots, ML, integrated platforms). Firm searches for continuous ways of improvement |
| | | occurs via internal paper courier services. | softwares (Slack, Skype) to handle operational purposes and get rid of paper-based systems. | Zoom, Microsoft Teams). | | towards its business operations. |
| | ICT-System Application | Firm's employees lack of knowledge as well as expertise to make use of ICT-Systems (e.g. computer programs, hardware, softwares). | Initiatives to set up ICT-Systems depend purely on few heroics employees' practices due to lack of know-how from the majority of employees. | Firm's employees act proactively and are actively encouraged through the management levels to learn and make use of ICT-Systems (e.g. computer programs, hardware, softwares). | ICT-Systems become tangible and well-integrated in the ICT-Infrastructure. Firm's employees are almost experts on handling such tools and thus, face little to almost no complications when using them. | ICT-Systems are fully implemented over the firm. Firm's employees become experts in ways that they are able to handle such tools without facing any complications. ICT-Systems are constantly evaluated and improved accordingly. |
| **Customer-Centricity** | Customer Satisfaction | Customers' satisfaction is not a priority for the firm, and management has little to no interest in finding ways to achieve it. | Customers' satisfaction is becoming important, but it depends, purely, on heroics front-line employees' practices. | Customers' satisfaction is recognized as important and it increases through the application of ICT-Systems (e.g. intelligent call routing, interactive voice responses, virtual receptionist) to support front-line employees' practices. | Customers' satisfaction is part of the firm's culture and it increases through the application of trend technologies (e.g. AI, AR, chatbots) to allow faster and more optimized solutions for customers' problems. | Customers are always satisfied because of the use of trend technologies (e.g. AI, AR, chatbots) and the firm's capacity to dispose of a digital platform, where customers and other users can co-create and assist in the creation of customers' solutions at all times. |
| | Customer Interaction | Customers' interaction rely, purely, on front-desk (e.g. face-to-face) as well as telephone encounters. | Customers' interaction relies, mainly, on front-desk as well as telephone encounters. Customers' interaction takes place by using only text-technologies | Customers' interaction is extended through the application of Internet-based applications (e.g. social media platforms), allowing customers and other users to | Customers' interaction increases through the application of trend technologies (e.g. real-time messaging, AI, chatbots) as well as by having the presence of | Customers' interaction is achieved at all scales through the presence of trend technologies (e.g. real-time messaging, AI, chatbots) as well as omni-channel |
| | | | | (e.g. email) as well as basic ICT-applications (e.g. internet website). | interact by creating, sharing, or exchanging information anytime. | omni-channel services. | services to optimize user experience. Firm disposes a digital platform for customers and other users to co-create and interact with each other, evading aspects such as time and space. |
| | Customers Empathy | Firm problems are only faced from the firm's perspective, ignoring the customer perspective. | Firm starts approaching the customer, and curiosity starts to raise, resulting in the firm's willingness to explore and discover the customer situation and experience. | Firm takes an active role and starts to wonder in the customer's world. Firm starts to acquire a customer-centered mindset by focusing on the customer as their biggest source of value. | Firm connects with the customer by recalling explicitly upon his/her own ideas, needs, and experiences to reflect and be able to understand what is it that the customer wants. Firm manages to connect on an emotional level with the customer by recalling upon his/her feelings and resonates with the customer's experience. | Firm creates an emotional connection with the customers and make sense, at any time, use of the customer's perspective. |
| **Strategy and Leadership** | Strategy and Vision | No formal strategy nor a clear and defined vision on how to assess issues related to digitalization. The urgency to transit from analog to digital for firm's own survival is not recognized. | Urgency of achieving a transition from analog to digital is not sufficiently recognized and consciously ignored. Development plans have, mostly, silos and static structures. | Urgency of achieving a transition from analog to digital is recognized as important. Business sees ICT as interdependent and acknowledges that business processes have to be revised to take advantages of ICT. | Urgency of achieving a transition from analog to digital is fully recognized at all levels of the firm. Firm relies on different types of ICT-Systems to support strategic goals. Firm's focus is aimed on achieving a business/IT alignment to assess efficiency | Strategy is fully supported with the implementation of ICT-Systems and it mostly aims to create competitive advantages and strategic differentiation. Business/IT alignment is fully achieved, and it shows positive results when handling organizational |

| | | | | | | and effectivity issues. | functions and processes. Digitalization becomes a central component of firm's vision, mission, and strategy. |
|---|---|---|---|---|---|---|---|
| | Business Model | Business model is either completely unknown or, predominantly, analog. Digitalization has no significance for the business idea. | Firm recognizes the importance of digitalization and makes use of few ICT tools to improve the business model. Firm lacks ICT resources. | First digital initiatives are launch through the use and application of ICT-Systems (e.g. enterprise resource management systems such as SAP or Oracle), social media platforms, and marketing channels to manage day-to-day business operations and activities. | Business model is based on the application and use of trend technologies (e.g. integrated platforms, AI, cloud-based services, chatbots) that enable the firm to achieve new ways of interaction as well as to optimize collaboration. | Business model is completely digitalized through the application and use of trend technologies (e.g. integrated platforms, AI, cloud-based services, chatbots). Firm constantly searches for ways to re-design and empower new forms of interaction and collaboration. |
| | Digital Culture | Firm neglects the importance of aspects related to innovation, collaboration, and openness. Firm considers that it can survive based on individual efforts. | Firm recognizes the importance of investing in ICT-Infrastructure, acquiring the respective licenses as well as necessary ICT-Systems to assess the firm's processes. | Firm starts to acquire a mindset based on inclusion and implements basic ICT-Systems to create digital solutions to expand the customer base, transform the customer experience, and achieve new forms of collaboration. | Digital culture is embedded in firm's strategy and mindset. Firm promotes the use and application of ICT-Systems at all levels of the firm. The implementation of ICT-Systems led to new forms of collaboration (e.g. co-creation) and interaction. | Firm encourages employees to look outward and engage with customers and partners through ICT-Systems to create new solutions (e.g. co-creation). Firm encourages boldness over caution. Employees are encouraged to take risks, fail fast, and learn. Firm supports the need for speed and promotes continuous iterations (e.g. prototyping) rather than perfecting a |

| | | | | | | | product or idea before launching it |
|---|---|---|---|---|---|---|---|
| | | | | | | | |

Table 5: Digital business ecosystem maturity model for personal service firms

<a>DISCUSSION

Recently, researchers such as Canetta et al. (2018), Jansen (2020), Azevedo and Santiago (2019) as well as Gollhardt et al. (2020) have presented systematic ways to achieve DBEs through the development of MMs. However, the application of such MMs is considered 'context-specific' (Canetta et al. 2018), meaning that they are hardly transferable and applicable in the context of PS firms. This led us to develop a new MM, which includes the requirements of PS firms. Based on our results, we found out that it is not only the use, application, and implementation of multiple technologies (e.g. AI, ML, integrated platforms, virtual agents, video-streaming softwares), which lead companies to achieve a DBE but rather the application of such technologies in combination with good management practices (e.g. collaboration, co-creation). These findings are in line with Teichert (2019), who argues that DBEs go beyond a 'mere' reflection of the extent to which a firm performs tasks and handles information flows by ICT but also requires a management and leadership performance to assess issues related to the company's strategy, collaboration, customer integration, mindset, and culture. Similarly, our results indicate how the implementation of ICT allows PS firms to find new settings to increase user's experiences and to improve aspects related to communication, interaction, collaboration, as well as co-creation, which are fundamental aspects for their DBE development. These findings are in line with authors such as Robra-Bissantz et al. (2020) and Lattemann et al. (2020), who emphasized that PS are all about the fulfillment of a user's needs (i.e. *Value-in-Use*) as well as the design of experiences for the user in interaction (i.e. *Value-in-Interaction*), whereby both could be highly influenced by ICT. Consequently, in line with Gunsberg et al. (2018), our results also show that by the application of agile methods (e.g. Design Thinking and SCRUM), firms are not only able to achieve new kinds of innovations with regards to product/services but also to improve DBEs in ways that companies enhance their capacity to collaborate and co-create with customers and other actors. Similarly, studies such as Vink (2018) have suggested different ways how firms can use social media networks (e.g. Facebook, Instagram), to enhance aspects related to 'customers empathy'. This is contrary to our results, where there was an agreement among the interviewees claiming that ICT could not enhance customers' empathy but rather discourage it.

Nowadays, to successfully ride the wave of change, PS firms need to evaluate how digital disruption is changing customer behavior continuously, rethink their business towards developing more individual- and customized solutions, and re-design customers' roles to form co-creation practices (Vargo and Lusch, 2008). Especially PS firms, which are services considered relying highly on high-contact levels of interaction, it is all about understanding customer requirements and preferences and delivering an outstanding experience at every customer touchpoint. Consequently, for DBEs to be a success, collaboration is crucial. While technology enables new ways of working, collaboration is the key catalyst for promoting the agile and management practices required to achieve DBEs (Camarinha-Matos et al. 2019). The need to share and combine information, knowledge, and other resources along the company as well as to develop agile coordination mechanisms to support efficiency in businesses processes, correspond to important facets of collaboration. Additionally, the emergence of strategic-long term partnerships such as DBEs require new organizational structures and advanced models of collaboration, in which the focus lies on the cooperation of different actors (i.e., service providers, customers) and stakeholders with the aim of applying collective knowledge to create value for the individual and the entire ecosystem (Akaka and Vargo 2015). Finally, digitalization is reducing demand for centralized and standardized routines, analogue and manual tasks, and siloes working styles while increasing demand for decentralized routines and virtual collaboration skills (Robra-Bissantz et al. 2020). Most PS firms start with a centralized model, where not only a founder or rather a manager makes all decisions but also where most of the company's processes are analogue (Bartik et al. 2020). As the business grow and diversify, their environments become more complex. PS firms need to become more flexible and responsive concerning the use of ICT, resulting in decentralized digital network structures, whereby collaboration and co-creation practices are executed among all company levels, also involving external partners. In this context, PS firms must question and re-evaluate traditional approaches to organize work and search for new organizational structures (e.g. DBEs) that can achieve efficiency but also have the flexibility for success in today's digital age.

<a>CONCLUSION

PS firms must recognize the advantages of ICT in ways that might lead these firms to improve information flows and aspects related to communication, interaction, collaboration, and co-creation, which are fundamental aspects when designing and developing DBE. Yet, the design of a DBE is complex, as it usually requires the adaptation of a certain type of ICT-infrastructure and the adaptation and conduction of vast management practices.

Our DBE MM for PS firms allows such complexity to be deconstructed in three ways. Firstly, it allows PS firms to identify what kind of technologies and management practices they need to apply and combine to achieve a DBE. Secondly, it also depicts the business functions, processes, or capability areas that need to be addressed by using such technologies and management practices. By doing this, our MM allows PS firms to identify room for improvements in their management as well as ICT-infrastructure. Thirdly, our MM might allow PS firms to develop more successful products/services and to be more customized and geared to customers' needs and desires. Consequently, we contribute to current literature and practice in several ways. First, we are the first to provide an 'industry-specific' and not 'context-specific' MM for PS firms by indicating several stages, dimensions, capability areas, and initiatives for improvements that describe the evolution path towards developing a DBE. Second, from an academic perspective, we respond to recent calls for more research on building a strategic instrument that allows PS firms to implement adequate processes and practices to effectively manage and guide the transition towards digital in a systematic way (Larsson, 2015). Third, from a practical perspective, our MM allows PS firms to improve their business model, functions, and processes as well as to assess their current 'digitalization' stage.

This research also has its limitations. First, a possible limitation regarding qualitative research is that it engages interviews to collect data and such could be susceptible to backwards reconstructions and false findings. Second, we are aware that our results were derived, including a limited number of companies' representatives as a data sample. However, to overcome these problems, we focused on company representatives based on their expertise and firsthand experience in PS firms, digitalization, and BE. Consequently, to address issues related to the reliability and applicability of our model as well as to offer more practical insights, our MM was later validated by two companies from our interview network. Finally, we further suggest exploring the issue of how the application and use of ICT might impact customers' empathy to see if there are other ways to refine our MM.

## ACKNOWLEDGEMENTS

This paper is part of the project "Begleitforschung Personennahe Dienstleistungen" (BeDien), funded by the German Federal Ministry of Education and Research (BMBF), grant numbers 02K17A080 and 81.